\documentclass[12pt]{article}
\usepackage{epsf}

\def\hybrid{\topmargin 0pt      \oddsidemargin 0pt
	\headheight 0pt \headsep 0pt
	\textwidth 6.25in       
        \textheight 9.5in       
	\marginparwidth .875in
	\parskip 5pt plus 1pt   \jot = 1.5ex}

\catcode`\@=11
\def\marginnote#1{}

\newcount\hour
\newcount\minute
\newtoks\amorpm
\hour=\time\divide\hour by60
\minute=\time{\multiply\hour by60 \global\advance\minute by-\hour}
\edef\standardtime{{\ifnum\hour<12 \global\amorpm={am}%
	\else\global\amorpm={pm}\advance\hour by-12 \fi
	\ifnum\hour=0 \hour=12 \fi
	\number\hour:\ifnum\minute<10 0\fi\number\minute\the\amorpm}}
\edef\militarytime{\number\hour:\ifnum\minute<10 0\fi\number\minute}

\def\draftlabel#1{{\@bsphack\if@filesw {\let\thepage\relax
   \xdef\@gtempa{\write\@auxout{\string
      \newlabel{#1}{{\@currentlabel}{\thepage}}}}}\@gtempa
   \if@nobreak \ifvmode\nobreak\fi\fi\fi\@esphack}
	\gdef\@eqnlabel{#1}}
\def\@eqnlabel{}
\def\@vacuum{}
\def\draftmarginnote#1{\marginpar{\raggedright\scriptsize\tt#1}}

\def\draft{\oddsidemargin -.5truein
	\def\@oddfoot{\sl preliminary draft \hfil
	\rm\thepage\hfil\sl\today\quad\militarytime}
	\let\@evenfoot\@oddfoot \overfullrule 3pt
	\let\label=\draftlabel
	\let\marginnote=\draftmarginnote
   \def\@eqnnum{(\theequation)\rlap{\kern\marginparsep\tt\@eqnlabel}%
\global\let\@eqnlabel\@vacuum}  }

\def\numberbysection{\@addtoreset{equation}{section}
	\def\theequation{\thesection.\arabic{equation}}}

\catcode`@=12
\relax

\def\nn{\nonumber}
 
\def\beq{\begin{equation}}
\def\eeq{\end{equation}}
\def\bea{\begin{eqnarray}}
\def\eea{\end{eqnarray}}

\relax
\numberbysection
\hybrid

\begin{document}
\begin{titlepage}
\begin{center}
{\large\bf Finite average lengths in critical loop models}\\[.3in] 
        {\bf Jesper Lykke Jacobsen and Jean Vannimenus} \\ 
	{\it Laboratoire de Physique Statistique%
             \footnote{Laboratoire associ{\'e} aux universit{\'e}s
                       Paris 6, Paris 7 et au CNRS.},\\
             Ecole Normale Sup{\'e}rieure,\\
             24 rue Lhomond,
             F-75231 Paris Cedex 05, France. \\}
\end{center}
\vskip .15in
\centerline{\bf ABSTRACT}
\begin{quotation}

{\small A relation between the average length of loops and their free
energy is obtained for a variety of O($n$)-type models on two-dimensional
lattices, by extending to finite temperatures a calculation due to Kast.
We show that the (number) averaged loop length $\overline{L}$ stays
finite for all non-zero fugacities $n$, and in particular it does
{\em not} diverge upon entering the critical regime ($n \to 2^+$).
Fully packed loop (FPL) models with $n=2$ seem to obey the simple
relation $\overline{L} = 3 L_{\rm min}$, where $L_{\rm min}$ is the
smallest loop length allowed by the underlying lattice. We demonstrate
this analytically for the FPL model on the honeycomb lattice and for
the 4-state Potts model on the square lattice, and based on numerical
estimates obtained from a transfer matrix method we conjecture that
this is also true for the two-flavour FPL model on the square lattice.
We present in addition numerical results for the average loop length
on the three critical branches (compact, dense and dilute) of the
O($n$) model on the honeycomb lattice, and discuss the limit $n \to 0$.
Contact is made with the predictions for the distribution of loop
lengths obtained by conformal invariance methods.}



\end{quotation}
\end{titlepage}

\section{Introduction}

Loop models appear in various contexts in statistical physics, not
only as models of various phases of polymers \cite{deGennes} but also
in the high-temperature expansion of the O($n$) spin model
\cite{Domany81,Nienhuis82}, in surface growth models \cite{surf_growth},
in efficient cluster-flipping  Monte Carlo methods \cite{Evertz}, and
in the study of quantum spin  systems \cite{Aizenman}. They have deep
connections with other well-studied statistical models, such as the
eight-vertex, colouring and Potts models \cite{Baxter82,jk_prb}, and
they can be viewed as toy models for the study of string theories.
One of their main
interests is that they provide examples of systems involving
{\em extended} degrees of freedom, rather than local ones as in spin
systems, for which many exact results have been obtained in $d=2$
using Bethe ansatz \cite{Baxter82}, Coulomb gas \cite{nien_rev} and
conformal invariance methods \cite{cardy_rev}.

The basic O($n$) loop model is defined through the partition function
\beq
 Z(n,t) = \sum_{\cal C} n^P t^V,
 \label{Z-On}
\eeq
where $n$ is the loop fugacity, which controls the number $P$ of
different loops (``paths'') in the configuration ${\cal C}$, and $t$
is a temperature-like variable which controls the number $V$ of empty
sites (``voids'') not covered by any loop, the total number of sites
on the lattice being $N$.%
\footnote{This is the standard definition of the O($n$) model on the
honeycomb latttice \cite{Nienhuis82}. On lattices with higher
coordination number a given site may be visited by more than one loop,
necessitating the introduction of further weights in Eq.~(\ref{Z-On})
\cite{Blote89}. We shall come back to this point later on.}
For non-crossing loops on the honeycomb and square lattices a phase
transition occurs at $n = n_{\rm c} = 2$, for all temperatures below a
lattice-dependent $t_{\rm c}$ \cite{Nienhuis82,Blote89}. That
transition is very weak and is reminiscent of the Kosterlitz-Thouless
transition for the XY spin model, as the free energy only has an
essential singularity. The geometrical interpretation is simple:
For $n > n_{\rm c}$ the configurations contributing to $Z$ only
contain finite loops in the thermodynamic limit, whilst for
$n \leq n_{\rm c}$ there appear ``infinite'' loops whose extent
diverges with the lattice size. This is analogous to what
happens in lattice percolation models, where an infinite cluster
appears when the site (or bond) occupation probability is above a
threshold value $p_{\rm c}$ \cite{Stauffer}, and indeed the perimeters
of percolation clusters can be viewed as special cases of loops
\cite{cardy_rev,Aizenman99}.

Explicit formulae for the free energy per site, defined as   
\beq
 F(n,t) = \lim_{N \to \infty} \frac{1}{N} \log Z(n,t),
 \label{F}
\eeq
can be obtained \cite{Baxter86,Batchelor88,Suzuki88} 
along certain critical lines in the $(n,t)$ plane \cite{Nienhuis82}
for the honeycomb lattice, but the distribution of loop lengths
cannot be inferred from the partition function, and information about
it is obtained numerically \cite{jk_prb} or through scaling and
conformal invariance methods \cite{DupSaleur87} (see the lectures
by Cardy \cite{cardy_rev} for an introduction to the geometric
properties of loop models). We show in the first part of the present
work that a simple relation, first derived by Kast at $t=0$ in the
honeycomb O($n$) model \cite{Kast}, can be generalised to finite
temperatures and to a variety of other loop models. It gives the
average loop length $\overline{L}$ from the knowledge of the free
energy and its derivatives with respect to $n$ and $t$ only.

In the second part we use the exactly known results for $F(n,t=0)$ to
calculate analytically that average length at the fugacity where a
transition occurs, $n_{\rm c} = 2$. We present next very accurate
numerical calculations of the average loop length, using a transfer matrix
method and our relation, at finite temperatures and for the
zero-temperature O($n$) model on the square lattice. For both the
$t=0$ (fully packed) model on the honeycomb lattice and for the
4-state Potts model on the square lattice we find the exact result
\beq
 \overline{L} = 3 L_{\rm min},
 \label{3}
\eeq
where $L_{\rm min}$ is the minimal loop length allowed by the given
lattice. Contrary to what was suggested in Ref.~\cite{Kast} on the
basis of numerical results, the average length thus remains {\em
finite} for $n=n_{\rm c}$, and indeed 
for all non-zero $n$. This is a priori counter-intuitive, as one would
naively expect the appearance of infinite loops below $n_{\rm c}$ to
be reflected in a diverging average length, but we show in the last
part of the paper that this result is not in contradiction with the
predictions of conformal invariance.

\section{Average loop length and the free energy}
\label{sec:av-length}

The average length of non-crossing loops in a given configuration
${\cal C}$ on the honeycomb lattice is given by 
\beq
 \langle L({\cal C}) \rangle =  \frac{N-V}{P},
 \label{L-honey}
\eeq
since a given site belongs to at most one loop, and its ensemble
average with respect to Eq.~(\ref{Z-On}) is 
\beq
 \overline{L} = \frac{1}{Z} \sum_{\cal C}
 \frac{N-V}{P} n^P t^V.
\eeq
Introducing the partial derivatives $\partial_n$ and $\partial_t$ with
respect to $n$ and $t$ one has:
\bea
 \partial_n (Z \overline{L}) &=&
 \frac{1}{n} \sum_{\cal C} (N-V) n^P t^V, \\
 \partial_t Z &=&
 \frac{1}{t} \sum_{\cal C} V n^P t^V.
\eea
Eliminating the average number of voids between these two relations gives
\beq
 \partial_n (Z \overline{L}) =
 \frac{1}{n} \left[ N Z - t \, \partial_t Z \right],
\eeq
or, in terms of the free energy,
\beq
 \frac{\partial_n \overline{L}}{N} +
 \overline{L} \, \partial_n F =
 \frac{1}{n} \left[ 1 - t \, \partial_t F \right].
 \label{deriv}
\eeq
Now, if $\partial_n \overline{L}$ remains finite the first term on the
left-hand side of Eq.~(\ref{deriv}) becomes 
negligible in the thermodynamic limit, and we obtain a simple relation
between  $\overline{L}$ and the partial derivatives of $F(n,t)$,
\beq
 \overline{L} = \frac{1 - t \, \partial_t F}
                     {n \, \partial_n F},
 \label{honey-Kast}
\eeq
which generalizes Kast's relation \cite{Kast} to non-zero temperature.  

Several remarks are in order here:
\begin{itemize}
\item The validity of relation (\ref{honey-Kast}) relies on the
 finiteness of $\partial_n \overline{L}$, which in turn,
 as is seen from Eq.~(\ref{honey-Kast}), holds as long as 
 $\partial_n \partial_t F$ and $\partial_n^2 F$
 remain finite, and $n$ and $\partial_n F$ do not vanish. As we
 shall see below, these conditions can be shown explicitly to be
 fulfilled for fully packed ($t=0$) loops on the honeycomb lattice, for
 all $n > 0$.
\item The knowledge of $F(n,t)$ along a critical line in the $(n,t)$
 plane is not sufficient to obtain the average length $\overline{L}$,
 as it provides only one relation between $\partial_n F$
 and $\partial_t F$, through the total derivative
 ${\rm d}_n F$, and in general it is not possible to express
 $\overline{L}$ as a function of ${\rm d}_n F$ alone. 
\item One can also write relations for $\overline{P}$, the ensemble
 average number of loops \cite{cardy_rev}:
 \beq
  \overline{P} = \frac{1}{Z} \sum_{\cal C} P({\cal C}) n^P t^V.
 \eeq
 One has from Eqs.~(\ref{Z-On}) and (\ref{F})
 \beq
  \overline{P} = \frac{n}{Z} \, \partial_n Z = N n \, \partial_n F.
  \label{P1}
 \eeq
 Taking now the derivative of Eq.~(\ref{P1}) with respect to $n$, one
 finds 
 \beq
  \partial_n \overline{P} =
  \frac{\partial_n Z}{Z} +
  n \left[ \frac{\partial_n^2 Z}{Z} -
  \left( \frac{\partial_n Z}{Z} \right)^2 \right] =
  \frac{1}{n} \left( \overline{P^2} - \overline{P}^2 \right),
 \eeq
 which yields an explicit expression for the fluctuations in the loop number:
 \beq
  \frac{1}{N} \left( \overline{P^2} - \overline{P}^2 \right) =
  n \, \partial_n F + n^2 \, \partial_n^2 F.
 \label{P2}
 \eeq
 Relations (\ref{P1}) and (\ref{P2}) show that the average number
 of loops is extensive (except for $n=0$) and that its fluctuations
 become negligible in the thermodynamic limit, as long as
 $\partial_n F$ and $\partial_n^2 F$ remain
 finite. It is also interesting to note that for a perfect gas at
 fixed chemical potential the fluctuations in the number $M$ of
 particles in a given volume are given by
 \beq
  \overline{M^2} - \overline{M}^2 = \overline{M},
 \eeq
 so the second term on the right-hand side of Eq.~(\ref{P2}) may be viewed
 as a correction with respect to the fluctuations in a ``perfect gas''
 of loops, induced by the non-overlapping condition.  
\end{itemize} 

Before leaving this Section we briefly comment on the appropriate form
of the relation (\ref{honey-Kast}) for lattices with higher
coordination number. As an example, consider the O($n$) model on the
square lattice, for which the partition function has a slightly more
complicated form than the one given in Eq.~(\ref{Z-On}),
since each vertex can be visited by the loops in several ways that are
unrelated by rotational symmetry. Following Ref.~\cite{Blote89} we define
\beq
 Z_{{\rm O}(n)} = \sum_{\cal C} t^{N_t} u^{N_u} v^{N_v} w^{N_w} n^P,
 \label{Z-square}
\eeq
where $N_t$, $N_u$, $N_v$ and $N_w$ are the number of vertices visited
by respectively zero, one turning, one straight, and two mutually
avoiding loop segments. The relation (\ref{L-honey}) must be replaced
by
\beq
 \langle L({\cal C}) \rangle = \frac{N-N_t+N_w}{P},
\eeq
and under assumptions on the higher derivatives identical to those
given above the generalisation of Eq.~(\ref{honey-Kast}) finally reads
\beq
 \overline{L} \, \partial_n F = \frac{1}{n}
 \left(1 - t \, \partial_t F + w \, \partial_w F \right).
\eeq

\section{Zero-temperature case: Fully packed loops}
\label{sec:honey}

At $t=0$ configurations containing voids do not contribute to $Z$,
and on the honeycomb lattice every site must belong to one and only
one loop. The system of loops is then called {\em fully packed} or
``compact'', to distinguish it from the ``dense'' phase
\cite{Nienhuis82,DupSaleur87} with a finite density of voids found at
non-zero $t$. A necessary condition for the $t=0$ line
to be stable (in the RG sense) is that $Z(n,t)$ is invariant under the
replacement of $t$ by $-t$ \cite{nien_fpl}.%
\footnote{Negative values of $t$ correspond to an antiferromagnetic
interaction in the spin language.}
This symmetry criterion is fulfilled because the honeycomb lattice is
bipartite and hence only allows for loops of {\em even} length. 
For the model at hand the condition is also sufficient (see
Ref.~\cite{fpl_48} for an example where this is not the case), and as
a consequence, the $t=0$ line is critical for $n \leq 2$ and fully
packed loops are in a different universality class than dense loops
\cite{nien_fpl,batch94,jk_jpa,jj_npb,jj_prl}.
The relation (\ref{honey-Kast}) reduces
for $t=0$ to the form obtained by Kast \cite{Kast}:
\beq
 \overline{L}(n) = \frac{1}{n F_0'(n)},
 \label{orig-Kast}
\eeq
where $F_0(n)=F(n,t=0)$, and it is particularly interesting as
explicit expressions are known 
for the free energy in that case, thanks to the work of Baxter
\cite{Baxter70}, Reshetikhin \cite{Reshetikhin}, Batchelor {\em et al.}
\cite{batch94}, and Kast himself \cite{Kast}. One has for the free
energy per site of the honeycomb lattice
\begin{itemize}
\item for $n < 2$:  
 \beq
  F_0(n) = \frac{1}{2} \int_{-\infty}^{\infty} {\rm d}x\,
  \frac{\sinh^2 \lambda x \, \sinh(\pi-\lambda)x}
       {x \, \sinh \pi x \, \sinh 3 \lambda x},
 \label{n<2}
 \eeq
 where $\lambda > 0$ and $n=2\cos \lambda$.
\item for $n > 2$:
 \beq
  F_0(n) = \frac{1}{2} \log \left[ q^{1/3} \prod_{m=1}^{\infty}
  \frac{(1-q^{-6m+2})^2}{(1-q^{-6m+4})(1-q^{-6m})} \right],
 \label{n>2}
 \eeq
 where $q = {\rm e}^\gamma$ and $n = 2 \cosh \gamma$.
\end{itemize}
Analytically it is known that at the critical point $n_{\rm c}=2$, $F_0(n)$
takes the value \cite{Baxter70,batch94}%
\footnote{Note that there is a misprint in Eq.~(7) of Ref.~\cite{batch94}.}
\bea
 F_0(2)&=&\int_0^{\infty} \frac{{\rm d}u}{u} \,
          \frac{\sinh^2 u}{\sinh 3u} \, {\rm e}^{-u}
        = - \frac{1}{2} \sum_{m=1}^{\infty}
          \log \left[ 1- \frac{1}{(3m-1)^2} \right] \nn \\
       &=&\frac{1}{2} \log \left[ \frac{3 \Gamma^3(1/3)}{4 \pi^2} \right]
        = 0.18956 \cdots.
\eea
The analytic calculation can be pushed further, noting that for $n<2$
Eq.~(\ref{n<2}) can be reexpressed as 
\bea
 F_0(\lambda) - F_0(\lambda=0) &=&
 \int_0^{\infty} {\rm d}u \, \frac{\sinh^3 u}{u \sinh 3u}
   \left[1 - \frac{1}{\tanh(\pi u/\lambda)} \right] \nn \\ 
 &=& - \int_0^{\infty} \frac{{\rm d}v}{v} \,
   \frac{\sinh^3(\lambda v/\pi)}{\sinh(3\lambda v/\pi)}
   \left[\frac{1}{\tanh v} - 1 \right].
\eea
This expression can then be expanded in powers of $\lambda$. To lowest
order this reads:
\beq
 F_0(\lambda) - F_0(\lambda=0) \simeq
 - \frac{\lambda^2}{3 \pi^2} \int_0^{\infty} {\rm d}v \, v \,
   \frac{{\rm e}^{-v}}{\sinh v}
 = - \frac{2 \lambda^2}{3 \pi^2} \int_0^{\infty} {\rm d}v \,
   \frac{v}{{\rm e}^{2v}-1}
 = - \frac{\lambda^2}{36}.
\eeq
In terms of the loop fugacity $n$, this yields
\beq
 \left. \frac{{\rm d}F_0}{{\rm d}n} \right|_{n \to 2^-} =
 \left. \frac{{\rm d}F_0}{{\rm d}\lambda}
        \frac{{\rm d}\lambda}{{\rm d}n} \right|_{\lambda \to 0^+} =
 \frac{1}{36},
 \label{dF/dn}
\eeq
so we obtain an unexpectedly simple result for the average length of
the fully packed loops at the critical point  
\beq
 \overline{L}_0 (n=2) = 18.
 \label{18}
\eeq

A similar analysis can be performed for $n>2$, rewriting for example
Eq.~(\ref{n>2}) as
\beq
 F_0(n) - F_0(2) = \frac{\gamma}{6} - \frac12 \sum_{m=1}^{\infty} f_m,
 \label{f-sum}
\eeq
with
\beq
 f_m = \log \left[ \frac{g \big( 6m \gamma \big)
       g \big([6m-4] \gamma \big)}{g^2 \big( [6m-2] \gamma \big) } \right]
\eeq
and $g(x) = \frac{1}{x} \big(1-{\rm e}^{-x} \big)$.
The terms in Eq.~(\ref{f-sum}) have been grouped so as to make it more
convenient to use the Taylor-McLaurin summation formula  
\beq
 \sum_{m=1}^{\infty} f_m = \int_{0}^{\infty} {\rm d}y \, f(y) -
 \frac{1}{2} f(0) - \frac{1}{12} f'(0) + \frac{1}{720}f'''(0) + \cdots.
\eeq
The evaluation of Eq.~(\ref{f-sum}) for small $\gamma$ then yields for
$\left. {\rm d}F_0 / {\rm d}n \right|_{n \to 2^+}$ the same value as
Eq.~(\ref{dF/dn}). 

The calculation of ${\rm d}^2 F_0 / {\rm d}n^2$ proceeds along similar
lines, and the end result reads 
\beq
 \frac{{\rm d}^2 F_0}{{\rm d}n^2} = \frac{1}{1080} \mbox{ \rm for } n=2,
 \label{d2F/dn2}
\eeq
the same value being obtained on both sides of the transition (more
generally, one expects all derivatives of $F_0(n)$ to be continuous for
$n=2$). 
As $\lambda=0$ ($n=2$) is the only point where Eq.~(\ref{n<2}) may
have a singularity, we conclude that Kast's relation (\ref{honey-Kast})
remains valid in the whole region $n<2$, which was not a priori
obvious from his derivation.  

As a by-product one also obtains from Eqs.~(\ref{P1}) and (\ref{P2})
the average number of fully packed loops for $n=2$: 
\beq
 \left. \overline{P_0} \right|_{n=2} = \frac{N}{18}
\eeq
and its fluctuations:
\beq
 \left. \overline{P_0^2} - \overline{P_0}^2 \right|_{n=2} =
  N \left( \frac{1}{18} + \frac{1}{270} \right)
\eeq

Another interesting case is the limit of $n \to 0$. This is the
compact polymer (Hamiltonian cycle) limit, when a single loop covers
all the lattice sites (with adequate boundary conditions). The
derivative of Eq.~(\ref{n<2}) can now be evaluated directly:
\beq
 \left. \frac{{\rm d}F_0}{{\rm d}\lambda} \right|_{\lambda=\pi/2} =
 \frac{4}{\pi} \int_0^{\infty} {\rm d}u \,
 \frac{1 - \cosh 2u}{(1+2 \cosh 2u)^2} =
 \frac{1}{\pi} - \frac{2}{3 \sqrt{3}}.
\eeq
Thus
\beq
 \left. \frac{{\rm d}F_0}{{\rm d}n} \right|_{n=0} =
 \left. \frac{{\rm d}F_0}{{\rm d}\lambda} \, 
        \frac{{\rm d}\lambda}{{\rm d}n} \right|_{\lambda=\pi/2} =
 \frac{1}{3\sqrt{3}} - \frac{1}{2\pi},
 \label{1/30.03}
\eeq
and invoking once again Eq.~(\ref{orig-Kast}) the average loop length 
$\overline{L}_0(n)$ diverges as
\beq
 \overline{L_0}(n) = \frac{30.0344\cdots}{n} + {\cal O}(1).
 \label{30.03}
\eeq
One can also show that the second derivative
$F''(0) = -0.00684676\cdots$ is finite.

\section{Numerical results on the honeycomb lattice}
 
\subsection{Zero temperature}

We have checked numerically the analytical expressions given in the
literature for the free energy of the O($n$) model at $t=0$, using a
previously developed transfer matrix method \cite{nien_fpl} to compute
the partition function for strips of finite width, up to
$W_{\rm max} = 15$. The results can be further analysed by exploiting
that the dominant finite-size corrections are known from conformal
invariance arguments
\cite{bcn,affleck}
\beq
 F_0(W) = F_0(\infty) - \frac{\pi c(n)}{6 W^2} + \cdots,
 \label{cc}
\eeq
where $c(n)$ is the central charge of the model. The latter has been
inferred from the Bethe ansatz solution of the model \cite{batch94}
\beq
 c(n) = 2 - \frac{6 e^2}{1-e} \mbox{ \rm with } n = 2 \cos(\pi e).
 \label{c-honey}
\eeq
Carrying out the numerical differentiation on $F_0(\infty)$ obtained
from Eq.~(\ref{cc}), rather than on the $F_0(W)$ which are related to
the leading eigenvalue of the transfer matrix spectra, we were able to
obtain sufficient accuracy to calculate the derivative ${\rm d}F_0 /
{\rm d}n$, and hence $\overline{L_0}(n)$, using formula (\ref{honey-Kast}).

\begin{figure}
 \begin{center}
 \leavevmode
 \epsfysize=200pt{\epsffile{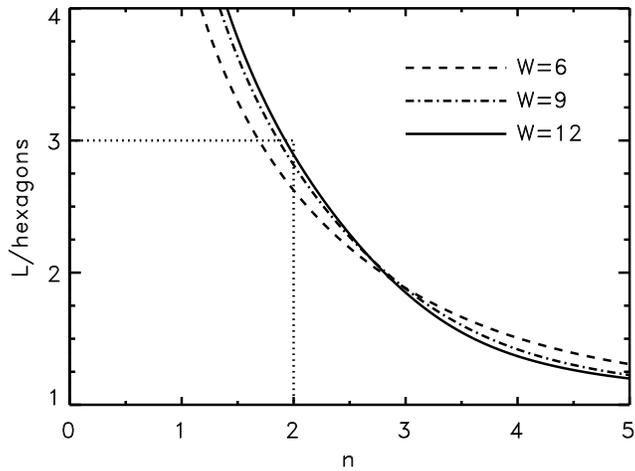}}
 \end{center}
 \protect\caption[2]{\label{fig:honey}Average loop lengths, measured
 in units of six, for the FPL model on the honeycomb lattice. The
 finite-size estimates converge rapidly towards the exact value
 $\overline{L}(n=2)=18$.}
\end{figure}

In Fig.~\ref{fig:honey} we plot $\overline{L_0}(n)$ in units of
$L_{\rm min}=6$, which is the shortest possible loop length on the
honeycomb lattice. To emphasize the fact that the average loop length
does not exhibit any singularity at $n_{\rm c}=2$ these data are shown
without the central charge correction. We point out that it would not
have been possible to produce this plot by a direct numerical
differentiation of the analytic results, Eqs.~(\ref{n<2}) and
(\ref{n>2}), as these expressions in their present form converge
extremely slowly close to the critical fugacity $n_{\rm c}=2$.
Actually we suspect that it was these difficulties that led Kast
\cite{Kast} to the erroneous conjecture that $\overline{L}(n \to 2^+)$
diverges.

\begin{table}
\begin{center}
\begin{tabular}{|r|l|} \hline
 $W$      & $F_0'(n=0)$ \\ \hline
   6      & 0.03192052  \\
   9      & 0.03304791  \\
  12      & 0.03321945  \\
  15      & 0.03326456  \\ \hline
 $\infty$ & 0.033296 (3) \\
 Eq.~(\ref{1/30.03}) & $0.0332951\cdots$ \\ \hline
\end{tabular}
\end{center}
\caption{\label{tab:honey0}Finite-size estimates of $F_0'(n=0)$,
corrected for the known 
value of $c$, as a function of strip width $W$. The extrapolated value
is found by fitting the residual size-dependence to a $1/W^4$ term.}
\end{table}

Turning now on the correction (\ref{cc}) we find the finite-size
values of $F_0'(n)$ given in Table~\ref{tab:honey0}. Results are only
shown for $W$ a multiple of three, since otherwise the equivalent
interfacial representation \cite{jk_jpa} will be subject to height
defects, leading to the introduction of non-trivial twist-like
operators in the continuum theory \cite{jj_npb}. An extrapolation to
the thermodynamic limit $W\to\infty$ can be performed by fitting the
residual size dependence to a $1/W^4$ term \cite{jj_npb}. The result
is
\beq
  \frac{1}{F_0'(n=0)} = 30.034 (3),
\eeq
in excellent agreement with the analytical result (\ref{30.03}). In
the same manner we find for $n=n_{\rm c}$ that
\beq
  \frac{1}{F_0'(n=2)} = 35.96 (3),
\eeq
confirming Eq.~(\ref{dF/dn}).

We have also evaluated numerically $F_0''(n=0) = -0.00686 (2)$, whereas
an attempt to compute $F_0''(n=2)$ did not lead to a sufficiently
accurate result to permit a comparison with Eq.~(\ref{d2F/dn2}). The
reason for this is a combination of the logarithmic corrections to the
scaling form (\ref{cc}) \cite{Cardy-log} and the singularity arising
in Eq.~(\ref{c-honey}) when taking the second derivative with respect
to $n$ at $n=n_{\rm c}$.

\subsection{Non-zero temperature: Dense and dilute loops}

As discussed earlier, explicit expressions for $F(n,t)$ are known
only along two other special critical lines in the $(n,t)$ plane,
for which Bethe ansatz solutions were given in
Ref.~\cite{Baxter86,Batchelor88,Suzuki88}.
These lines, or branches, are given by the relation \cite{Nienhuis82}
\beq
 t^2 = 2 \pm  \sqrt{2-n},
\eeq
where the upper sign corresponds to the ``dilute'' branch and the
lower sign to the ``dense'' branch. We have already remarked that this
knowledge is not sufficient to obtain analytically the average loop
length from Eq.~(\ref{honey-Kast}). However, an accurate numerical
determination of the partial derivatives $\partial_n F$
and $\partial_t Ft$ is possible, using the same transfer
matrix approach as for $t=0$ so the variation of $\overline{L}$ along
these lines can be studied in detail. The central charge values used
in the extrapolation procedure are now
\beq
 c(n) = 1 - \frac{6 e^2}{1 \mp e} \mbox{ \rm with } n = 2 \cos(\pi e).
 \label{c-dil-den}
\eeq

\begin{figure}
 \begin{center}
 \leavevmode
 \epsfysize=200pt{\epsffile{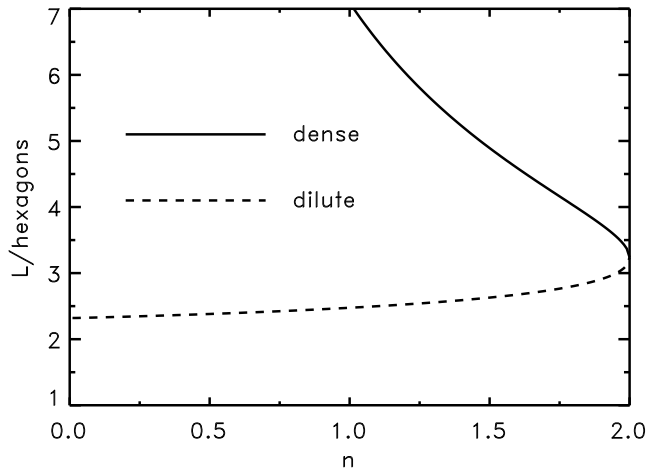}}
 \end{center}
 \protect\caption[2]{\label{fig:dil-den}Average lengths of dense and dilute
 loops on the honeycomb lattice, measured in units of six.}
\end{figure}

The resulting values of $\overline{L}(n)$ are displayed in
Fig.~\ref{fig:dil-den}. Only the data for strip width $W=9$ are
displayed, since employing the correction (\ref{c-dil-den}) renders the
residual finite-size effects indistinguishable on the figure.
Extrapolating to $W=\infty$ we obtain for $n = n_{\rm c} = 2$
\beq
 \overline{L}(n_{\rm c}, t_{\rm c}= \sqrt{2}) = 19.417 (1),
\eeq
which is slightly larger than the corresponding value at $t=0$. The
same is true for the asymptotic $n \to 0$ behaviour on the dense
branch, for which the numeric result is
\beq
 \overline{L}_{\rm dense}(n) = \frac{35.70 (2)}{n} + {\cal O}(1).
\eeq

At first sight this finding may seem odd, since the geometrical
scaling dimension
\beq
 x_2 = \frac12 (1-e) - \frac{e^2}{2(1-e)}
 \label{x2comp}
\eeq
coincides for the dense and the compact branch
\cite{nien_rev,batch94}, implying that the asymptotic distributions of
large loops are identical in the two models \cite{DupSaleur87}.
However the dense phase has fewer {\em short} loops than the compact
one, since it is always energetically favourable to trade a length-six
loop for some empty space:
\beq
 \frac{t^6}{n} = \frac{(2-\sqrt{2-n})^3}{n} \geq 1
 \mbox{ (\rm identity at $n=1$)}.
\eeq
Since a length-six loop necessarily has a fixed shape of its
perimeter there is no extensive entropy involved in the argument, and
our simple energetical consideration suffices.

Another remark pertains to the fact that $\overline{L}(n)$ is an
{\em increasing} function of $n$ on the dilute branch (and in
particular it does not diverge as $n\to 0$). This has a quite
elementary explanation in terms of $x_2$ to which we shall come back
in Section \ref{sec:CFT}.

\section{Loops on the square lattice}

Many results have also been obtained for systems of loops on the
square lattice \cite{Baxter82,jk_prb,Blote89,jj_npb,jj_prl,
Batchelor89,jk_npb,batch96,jk_prl,jj_jsp},
but now there are several possible definitions of the models,
depending on whether one allows the non-overlapping loops to cross at
the vertices or not, if there is only one kind of loops or two, and
exactly how the loop segments are allowed to turn at the vertices.
Following  Refs.~\cite{jj_npb,jj_prl}, we consider a model with two
kinds (``flavours'') of loops, with respective fugacities $n_1$ and
$n_2$, which both cover all $N$ vertices of the square lattice and can
intersect but not overlap. This can be considered as a particular
zero-temperature limit \cite{jj_jsp} of the O($n$) model on the square
lattice \cite{Blote89}. The partition function and the free energy
are given by   
\bea
 Z_0(n_1,n_2) &=& \sum_{\cal C} n_1^{P_1} n_2^{P_2},
 \label{FPL2} \\
 F_0(n_1,n_2) &=& \lim_{N\to\infty} \frac{1}{N} \log Z_0(n_1,n_2),
\eea
where the summation is performed over all doubly compact
configurations. The constraint that both types of loops cover all $N$
sites implies that for each configuration the average length of type-1
loops is just $\langle L_1({\cal C}) \rangle = N/P_1$,
and its ensemble average is given by
\beq
 \overline{L_1} =  \frac{1}{Z_0} \sum_{\cal C}
 \frac{N}{P_1} n_1^{P_1} n_2^{P_2}.
\eeq
Repeating the calculation of Section \ref{sec:av-length} one obtains
\beq
 \partial_{n_1} (Z_0 \overline{L_1}) = 
   N \frac{Z_0}{n_1}
\eeq
and
\beq
 \frac{1}{N} \partial_{n_1} \overline{L_1} + 
 \overline{L_1} \partial_{n_1} F_0 = \frac{1}{n_1}.
 \label{2flav}
\eeq
Now, if $\partial_{n_1} \overline{L_1}$ remains finite, the
first term in the left-hand side of Eq.~(\ref{2flav}) is negligible in
the thermodynamic limit, and we obtain  
\beq
 \overline{L_1} = \frac{1}{n_1 \partial_{n_1} F_0}.
 \label{FPL2-Kast}
\eeq
A similar relation holds for the average length $\overline{L_2}$ of
the second kind of loops, with $n_2$ replacing $n_1$, and when
$n_1 = n_2$ both reduce to Kast's relation (\ref{orig-Kast}).

One can also define the average length $\overline{L}$ of {\em any}
loop, regardless of flavour. Following the above line of reasoning this
is easily shown to fulfill the relation
\beq
 \overline{L} = 2 \frac{\overline{L_1} \, \overline{L_2}}
                       {\overline{L_1} + \overline{L_2}}.
\eeq
For $n_1 = n_2$ the above statement is trivial. Taking however the limit
$\overline{L_2} \to \infty$ we obtain the curious result that
\beq
 \overline{L} = 2 \overline{L_1} \mbox{ \rm for } n_2 \to 0,
\eeq
regardless of the value of $n_1$.

Since the model (\ref{FPL2}) has as yet defied a Bethe ansatz
solution, no analytical expressions for $F_0(n_1,n_2)$ are available
(apart from the conjecture $F_0(0,0) = \frac12 \log 2$ \cite{jj_npb}).
We have therefore studied numerically the relation (\ref{FPL2-Kast})
as a function of $n_1$, for different values of $n_2 = 0, 1$ and $2$.
Transfer matrix computations were performed for strips of even width,
up to $W_{\rm max} = 12$, exploiting that \cite{jj_npb}
\beq
 c(n_1,n_2) = 3 - \frac{6 e_1^2}{1-e_1} - \frac{6 e_2^2}{1-e_2}.
\eeq

Extrapolating as usual to $W=\infty$ we find for the average loop
lengths at $n_1 = 2$:
\beq
 \overline{L_1}(n_1=2,n_2) = \left \lbrace
 \begin{array}{lll}
   15.74 (5) \mbox{ \rm for } n_2 = 0 \\
   13.95 (3) \mbox{ \rm for } n_2 = 1 \\
   11.99 (2) \mbox{ \rm for } n_2 = 2 \\
 \end{array} \right.
\eeq
and for its divergence when $n_1 \to 0$:
\beq
 \overline{L_1}(n_1\to 0,n_2) = \frac{A}{n_1} + {\cal O}(1)
 \mbox{ \rm with } \left \lbrace
 \begin{array}{lll}
   A = 21.97 (6) \mbox{ \rm for } n_2 = 0 \\
   A = 20.35 (3) \mbox{ \rm for } n_2 = 1 \\
   A = 18.96 (3) \mbox{ \rm for } n_2 = 2 \\
 \end{array} \right. .
\eeq
A striking fact is that for the symmetric case $n_1 = n_2 = n_{\rm c} = 2$,
which corresponds to the four-colouring model introduced in
Ref.~\cite{jk_prb}, the numerical value of the average length of either
kind of loops is
\beq
 \overline{L_{\rm c}} / L_{\rm min} =  2.997 (4),
 \label{almost-3}
\eeq
where we have normalised with respect to the length of the shortest
possible loop on the square lattice, $L_{\rm min}=4$. The ratio
(\ref{almost-3}) is extremely close to 3, which by Eq.~(\ref{18}) is
the exact result for the zero-temperature O($n$) model on the
{\em honeycomb} lattice.

In order to see if this could be more than a coincidence, we have
looked at another model on the square lattice, describing the
$q$-state Potts model at its selfdual point \cite{Baxter82}.
We briefly recall how the Potts model is transformed into a loop model
\cite{Baxter82}: In a first step the spin model (with coupling
constant $K$) is turned into a random cluster model
\cite{Kasteleyn63}. Each state is now a bond percolation graph in
which each connected cluster is weighed by $q$ and each coloured bond
carries a factor of $({\rm e}^K-1)$. Next, the clusters are traded for
loops on the medial graph (the lattice formed by the mid-points
of the bonds on the original lattice). By the Euler relation each
closed loop is weighed by $\sqrt{q}$, and the bond weights are
$({\rm e}^K-1)/\sqrt{q}$ \cite{Baxter82}. At the selfdual point the
latter equals one by duality, and the exact correspondence is
\beq
  Z_{\rm Potts} = q^{N/2} \sum_{\cal C} \sqrt{q}^P
                \equiv q^{N/2} Z_{\rm Loop},
\eeq
where $N$ is the number of sites of the original lattice.

Defining $F$ as the free energy per medial site with respect to
$Z_{\rm Loop}$, one has \cite{Baxter82}
\begin{itemize}
 \item for $n<2$:
  \beq
   F(n) = \frac12 \int_{-\infty}^{\infty} {\rm d}x \,
   \frac{\sinh \lambda x \, \sinh(\pi - \lambda)x}
        {x \, \sinh \pi x \, \cosh \lambda x},
  \eeq
  where $\lambda > 0$ and $n = \sqrt{q} = 2 \cos \lambda$.
 \item for $n>2$:
  \beq
   F(n) = \frac{\lambda}{2} +
   \sum_{m=1}^{\infty} \frac{{\rm e}^{-m \lambda}}{m} \, \tanh m \lambda,
  \eeq
  where $n = \sqrt{q} = 2 \cosh \lambda$.
\end{itemize}
The value at the critical point $n_{\rm c}=2$ is also known:
\beq
 F(n=2) = \int_0^{\infty} \frac{{\rm d}u}{u} \, \tanh u \, {\rm e}^{-u}
        = 0.78318 \cdots,
\eeq
and as in Section \ref{sec:honey} we can evaluate the derivative
$\left. {\rm d}F/{\rm d}n \right|_{n \to 2^-}$ by examining the
difference
\beq
 F(\lambda) - F(\lambda=0) =
 - \int_0^{\infty} \frac{{\rm d}v}{v} \sinh (\lambda v/\pi)
 \tanh(\lambda v/\pi) \left[ \frac{1}{\tanh v} - 1 \right].
\eeq
At the bottom line we obtain
\beq
 \overline{L}(n=2) = 12,
\eeq
or $\overline{L_{\rm c}} / L_{\rm min} = 3$, the same result being
found also for $n \to 2^+$.
For $n \to 0$ the average loop length diverges as
\beq
 \overline{L}(n) = \frac{16}{n} + {\cal O}(1).
\eeq
Both these results are in excellent agreement with numerical estimates
produced by the transfer matrices described in Ref.~\cite{3-Potts}.

Our computations for the Potts model suggest that the result (\ref{3})
is not a mere coincidence, and we feel confident in conjecturing that
$\overline{L}/L_{\rm min} = 3$ is also exactly true for the
four-colouring model, cf.~Eq.~(\ref{almost-3}).
It would be quite surprising if this ratio were a new universal
quantity, like for instance the ratio of the average loop area to the
average square gyration radius, calculated by Cardy \cite{cardy_rev}.
Rather it is reminiscent of the ``quasi-invariants'' found in
percolation theory: For a large variety of lattices, the percolation
thresholds are observed to be little lattice-dependent, for a fixed
dimensionality, once they are properly normalised taking into account
the coordination number \cite{quasi-inv}.

\section{Discussion and contact with conformal invariance predictions}
\label{sec:CFT}

The most striking aspect of the results presented above is that the
average loop length remains finite, even inside the critical phase
where the correlation length, which is roughly the size of the
largest loop in a typical configuration, diverges.
In fact, a little thought shows that the
apparent paradox comes from the definition of ``average length'':
In our calculations all loops are equally weighted,
irrespective of their length. This is called the ``number average''
in polymer theory, to distinguish it from the ``weight average'',
where each polymer contributes with a weight proportional to its
length. Depending on the quantity being measured, these and
still other averages are relevant in order to compare
theoretical predictions with experiments (see, e.g., the
treatise by des Cloizeaux and Jannink \cite{Cloizeaux} for a detailed
discussion).
An analogous situation arises in percolation theory, when one defines
the ``average cluster size'' \cite{Stauffer}: If one means ``the average
size of the cluster on which an `ant' lands at random'', the weight
average has to be taken, which is different from the average size
obtained by simply giving every cluster the same weight. 
In practice these various averages may be very different when the
quantity considered has a power law distribution, as the result may
be dominated in one case by the few largest polymers or clusters,
and in another case by the numerous small ones. This is precisely
what happens in the loop systems considered here: In the critical
phase ($n \leq 2$) the (ensemble averaged) probability distribution of
loop lengths (unweighted) is expected to be of the form 
\beq
 \Pi(L) \sim L^{-\tau},
 \label{Pi}
\eeq
where the exponent $\tau$ is related to the geometrical (string)
scaling dimension \cite{DupSaleur87,surf_growth} $x_2$ by
\beq
 \tau - 1 = \frac{2}{2-x_2}.
 \label{tau}
\eeq
Conformal invariance arguments have permitted the exact evaluation of
$x_2$ for all loop models discussed in this paper. These are
\cite{nien_rev,DupSaleur87,batch94,jj_npb,jj_prl}
\beq
 x_2 = \frac{1}{2} (1 \mp e) - \frac{e^2}{2(1 \mp e)}
 \mbox{ \rm with } n = 2 \cos(\pi e),
 \label{x2}
\eeq
where the upper sign holds for any one of the compact or dense phases,
and the lower sign for the dilute O($n$) phase (see Ref.~\cite{jj_jsp}
for a discussion of this kind of universality). In particular
$3 > \tau >  2$ for all $0 < n < 2$. This implies that the
{\em number} averaged loop length 
\beq
 \overline{L} = \left( \sum_{L=L_{\rm min}}^{L_{\rm max}} L \Pi(L) \right) /
                \left( \sum_{L=L_{\rm min}}^{L_{\rm max}} \Pi(L) \right)
\eeq
converges, as we have shown explicitly. On the other hand, one can
pick  a point  at random on the lattice and ask what is the average
length of the loop to which it belongs. That {\em weight} averaged
length is given by 
\beq
 L^*= \left( \sum_{L=L_{\rm min}}^{L_{\rm max}} L^2 \Pi(L) \right) /
      \left( \sum_{L=L_{\rm min}}^{L_{\rm max}} L \Pi(L) \right)
\eeq
and it diverges like $L_{\rm max}^{3-\tau}$ for large system sizes.
In these equations $L_{\rm max}$ is {\em not} the largest loop that
can be placed on the $N$-site lattice (a Hamiltonian cycle of length
$N$), but rather loop size up to which the scaling law (\ref{Pi}) is
valid. On an $\ell \times \ell$ lattice this is given by
$L_{\rm max} \sim \ell^{D_{\rm f}}$, where \cite{DupSaleur87}
\beq
 D_{\rm f} = 2 - x_2
\eeq
is the fractal dimension of the loop. Using Eq.~(\ref{tau}) we then
find
\beq
  L^* \sim (\ell^2)^{D_{\rm f}-1} \sim N^{D_{\rm f}-1}.
 \label{FSS}
\eeq

\begin{figure}
 \begin{center}
 \leavevmode
 \epsfysize=200pt{\epsffile{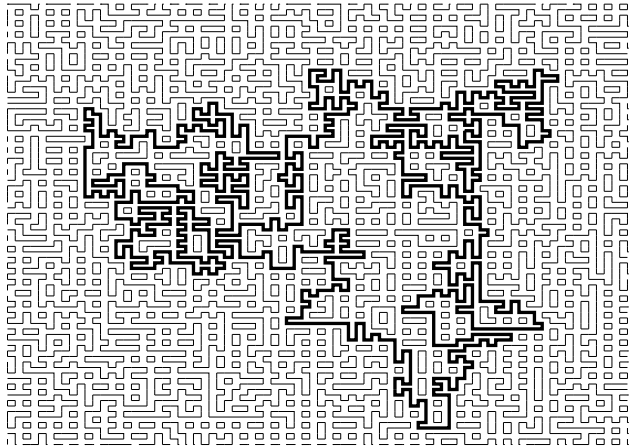}}
 \end{center}
 \protect\caption[2]{\label{fig:config}Detail of a typical
 configuration in the four-colouring model. The two loop flavours
 consist of the black and white lattice edges respectively. A loop
 passing through a randomly chosen point is likely to be very long,
 like the one of length 1076 shown in boldface. The presence of
 numerous short loops, however, explains why the (number) averaged
 loop length is finite in the thermodynamic limit: $\overline{L} = 12$.}
\end{figure}

For the four-colouring model Kondev and Henley \cite{jk_prb} have
confirmed this picture in detail using Monte Carlo simulations with
non-local loop updates. In particular the power-law distribution of
loop lengths (\ref{Pi}) was confirmed very accurately, with the
expected value $\tau = 7/3$. However, these authors did not examine
the finite-size dependence of $L^*$. In order to verify the validity
of Eq.~(\ref{FSS}) we have performed Monte Carlo simulations along the
lines of Ref.~\cite{jk_prb} on several lattice sizes, up to
$\ell_{\rm max}=200$. In each case the system was equilibrated by
means of $10^5$ loop flips, whereafter the lengths of a further
$10^6$ loop updates were registered. (A detail of a typical loop
configuration is shown in Fig.~\ref{fig:config}.) We found that
\beq
 L^* \sim \ell^{1.03 \pm 0.04},
\eeq
in good agreement with Eq.~(\ref{FSS}), since the fractal dimension is
known to be exactly $D_{\rm f} = 3/2$ \cite{jk_prb}.
Unfortunately, it is not possible to
relate $L^*$ directly to the partition function and to obtain
relations analogous to those holding for $\overline{L}$.

Let us end by commenting on the divergence of $\overline{L}$ as $n\to 0$.
By naively integrating Eq.~(\ref{Pi}) with a lower cut-off $L_{\rm min}$ 
and using scaling relations one finds
\beq
 \overline{L}(n) \sim \frac{2 L_{\rm min}}{x_2}.
\eeq
Since from Eq.~(\ref{x2comp}) $x_2 \to 0$ as $n \to 0$, it is not
surprising that $\overline{L}$ is found to diverge for compact and
dense loops. On the other hand, for the dilute phase $x_2$ as given by
Eq.~(\ref{x2}) does not vanish as $n\to 0$, in accordance with
Fig.~\ref{fig:dil-den}. In fact one has the rough estimate
\beq
 \frac{\overline{L}(n=2)}{\overline{L}(n=0)} \simeq \frac{3}{4}
\eeq
which is not too far off the numerical value of $\approx 0.72$.

\noindent{\large\bf Acknowledgments}

We thank  B. Derrida, E. Domany, V. Hakim, J. Kondev and especially
B. Nienhuis for discussions and encouragements during the course of
this work. JLJ is supported by CNRS through a position as
{\em chercheur associ\'e}.

\end{document}